\documentclass{article}

\usepackage[a4paper, total={6in, 8in}]{geometry}

\usepackage{titling}

\usepackage[most]{tcolorbox}

\newtcolorbox{greybox}{
  colback=gray!10,
  colframe=gray!50,
  boxrule=0.4pt,
  arc=1mm,
  left=7pt,
  right=7pt,
  top=7pt,
  bottom=7pt
}

\usepackage{lipsum}
\usepackage{endnotes}
\usepackage{natbib}
\usepackage{graphicx}
\usepackage{moreverb,url}

\usepackage{lipsum}  
\usepackage{footmisc}
\usepackage{enumitem}
\usepackage{float}

\usepackage{amsmath}
\usepackage{amssymb}
\usepackage{authblk}
\usepackage{tabularx}
\usepackage{array}

\usepackage{etoolbox}

\makeatletter
\patchcmd{\abstract}
  {\small}
  {\normalsize}
  {}{}
\makeatother

\usepackage{csquotes}

\usepackage[latin1]{inputenc}

\usepackage{musicography} 

\newcommand{\citeauthorpos}[1]{{\citeauthor{#1}'s}}

\newcommand{\citepos}[1]{{\citeauthorpos{#1}~\citeyearpar{#1}}}

\title{An introduction to pitch strength in contemporary popular music analysis and production \\
\ \\ 
\small In Music 2024, Innovation in Music Conference \\
 14-16 June, 2024, Kristiania University College, Oslo, Norway}
 
\author[1,2]{\small{Emmanuel Deruty}}
\affil[1]{Sony Computer Science Laboratories, 6 rue Amyot, 75005 Paris, France}
\affil[2]{Department of Architecture, Design and Media Technology, Aalborg University, Aalborg, Denmark}

\date{}

\begin{document}
\maketitle

\begin{greybox}
Published version: Deruty, Emmanuel (2026). An introduction to pitch strength in contemporary popular music analysis and production. In: Paterson, Justin, Rob Toulson, Russell Hepworth-Sawyer, Jan-Olof Gull\"{o}, Mark Marrington, and Claus Sohn Andersen, \textit{Innovation in music: current research perspectives}. \url{https://doi.org/10.4324/9781003475675}
\end{greybox}

\vspace{2cm}

\begin{abstract}

Music information retrieval distinguishes between low- and high-level descriptions of music. Current generative AI models rely on text descriptions that are higher level than the controls familiar to studio musicians. Pitch strength, a low-level perceptual parameter of contemporary popular music, may be one feature that could make such AI models more suited to music production. Signal and perceptual analyses suggest that pitch strength (1) varies significantly across and inside songs; (2) contributes to both small- and large-scale structure; (3) contributes to the handling of polyphonic dissonance; and (4) may be a feature of upper harmonics made audible in a perspective of perceptual richness.

\end{abstract}


\newpage
\section{Introduction: are text-to-music models suited to music production?}\label{sec:introduction}

\paragraph{Introduction.} The description of contemporary popular music has relied on a blend of high-level descriptors and low-level perceptual parameters. While high-level text descriptors, such as those used in generative AI models like MusicLM, offer a broad understanding of musical characteristics, they may not suffice for music production. This chapter examines the concept of pitch strength (PS) -- a low-level perceptual parameter -- and its relevance in this context. By exploring its variability across songs, impact on musical structure, role in managing polyphonic dissonance, and its connection to audible harmonics, we propose that integrating PS into generative AI frameworks could contribute to bridging the gap between machine learning models and music production practices.

The chapter is structured as follows: it begins with an examination of low- and high-level features in generative AI models, followed by a detailed characterization of PS. Next, the variability of PS in contemporary popular music is explored. Subsequent sections delve into the relationships between PS and various musical aspects, such as structure, dissonance, and perceptual richness. The chapter concludes with appendices that provide additional details and technical insights into specific aspects of the study.

\paragraph{High-level text captions in generative AI.} Music information retrieval has distinguished between quantifiable low-level audio attributes and informative high-level semantic descriptors \citep{bogdanov2009similarity,mckay2004automatic,zanoni2012features}.

The user input of recent generative text-to-audio models, such as MusicLM \citep{agostinelli2023musiclm}, consists of text captions. MusicLM's training involves text-audio pairs from MuLan \citep{huang2022mulan} and MusicCaps \citep{agostinelli2023musiclm}. MuLan uses annotations mapped to 44 million internet music videos. MusicCaps contains 5,521 music examples labeled by musicians. Using both MuLan and MusicCaps, MusicLM generates content from captions like: `Slow tempo, bass-and-drums-led reggae song. Sustained electric guitar. High-pitched bongos with ringing tones. Vocals are relaxed with a laid-back feel, very expressive'.

Such captions may be too high-level for the production of contemporary popular music, as defined by \citet{deruty2022development}. As noted by \citet[p.~8]{deruty2022melatonin} and exemplified by software synthesizers like Spectrasonics Omnisphere \citep{nagle2015omnisphere} or the extensive guitar setups of artists like My Bloody Valentine's Kevin Shields \citep{brewster2024shields}, low-level controls are indispensable to studio musicians.

\begin{figure}[htbp]
  \centering
  \includegraphics[width=1\columnwidth]{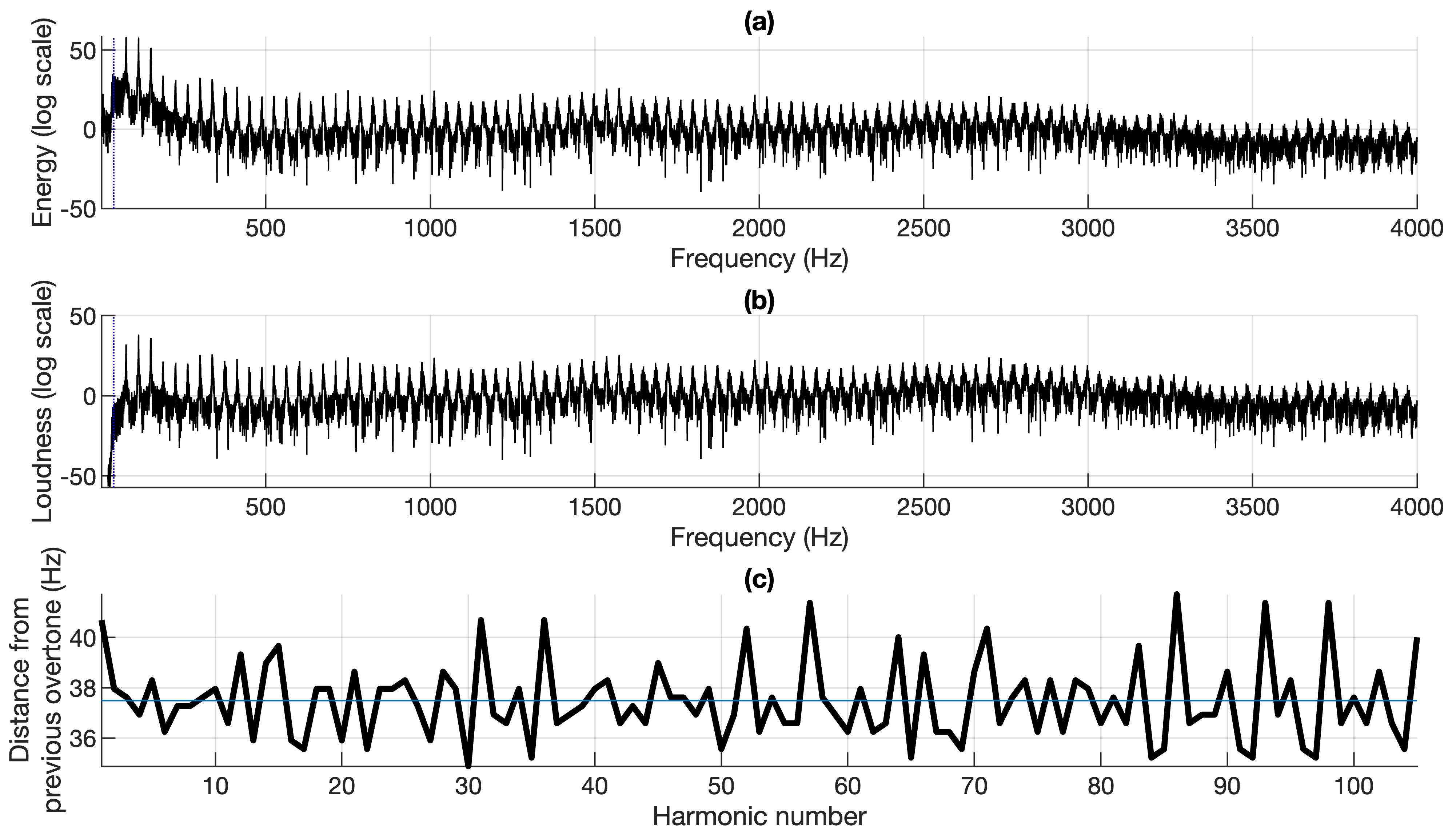}
  \caption{End guitar chord from \citet{mybloodyvalentine1991sometimes}. (a) Power spectrum; (b) weighted power spectrum (see Appendix \ref{ref:appendix1}); (c) frequency difference between consecutive overtones.}
\label{fig:sometimes}
\end{figure}

\paragraph{Example of low-level text description.} My Bloody Valentine's 1991 album Loveless was deemed influential and inspired several other artists \citep{hudson2021loveless}. \textbf{Figure~\ref{fig:sometimes}} shows an analysis of a guitar chord from \citet{mybloodyvalentine1991sometimes}. A possible text caption for this chord might be: `A 38Hz quasi-harmonic tone with no fundamental, high upper partials, random deviation of partials from harmonicity, and noise above 3000Hz'. As explained later, the caption may also involve pitch strength.

Such a caption is much lower level than those used in MusicLM. Musicians may find no interest in a guitar generated by recent text-to-audio models, as they can't fine-tune the guitar sound. From this perspective, generative AI models could benefit from incorporating lower-level features.

\emph{Pitch strength} (PS), a \emph{perceptual parameter} in the sense of \citet{mcadams1999perspectives}, is an example of a low-level aspect of contemporary popular music that may be useful in this context.

\vspace{1cm}

\paragraph{Chapter plan.} The chapter is organized as follows: first, a characterization of PS is provided. While the literature suggests that PS can be measured using the MPEG7 feature HarmonicRatio, this claim is shown not to be valid in the general case. Then, examples of variable PS in contemporary popular music are given. In particular, it is demonstrated that contemporary popular music covers a wide range of PS. Finally, relations between PS and musical structure, dissonance, and audible higher harmonics are explored. 

\vspace{1cm}

\newpage

\section{Pitch strength, characterization}\label{sec:characterization}

\paragraph{Definition.} \citet[p.~1708]{yost2009pitch} defines pitch strength as:

\begin{quote}
`\emph{Pitch strength} (also called \emph{pitch saliency}) refers to the strength of the perceived pitch of a complex sound as compared with the overall perceived quality (timbre) of the sound.'
\end{quote}

\paragraph{Signal features and pitch strength.} \citet{zwicker1990pitchstrength} report that:

\begin{itemize}[noitemsep]
	\item Sounds with line spectra generate a strong pitch, while continuous spectra produce a weaker pitch, except for narrow-band noise.
	\item PS of pure tones increases with duration and sound pressure, peaking at mid-range frequencies.
	\item Expanded bandwidth of noise bands reduces PS.
	\item Peaked ripple noise can achieve PS similar to complex tones.
	\item Low-pass noise with steep spectral slopes has significantly lower PS than pure tones, and the PS decreases further with less steep slopes.
	\item Partial masking sounds can significantly reduce the PS of pure tones.
\end{itemize}

\citet[pp.~412--413]{wightman1973pattern} add:

\begin{itemize}[noitemsep]
	\item More harmonics result in a stronger pitch.
	\item Harmonic complex tones have a stronger pitch than inharmonic ones.
\end{itemize}

\paragraph{Pitch strength and iterative rippled noise.} PS has been extensively studied in iterative rippled noise (IRN) \citep{patterson1996relative,patterson2000perceptual,yost1996pitch,yost1978strength,yost1978broadband,yost1994testing,yost1996timedomain}. According to \citet{patterson1996relative,yost1994testing,yost1996timedomain}, the PS of an IRN wave is related to the height of the first peak in its normalized autocorrelation function (AC1), equivalent to the HarmonicRatio MPEG7 feature \citep{isompeg7audio}. Specifically, the PS of an IRN wave is an exponential function of AC1: $PS = k 10^{AC1}$, where $k$ is a proportionality constant.

\paragraph{Pitch strength and HarmonicRatio.} HarmonicRatio (AC1) has been described as `the proportion of harmonic components within the power spectrum' \citep[p.~36]{isompeg7audio} and `the ratio of harmonic power to total power' \citep[p.~33]{moreau2006mpeg}. Given that it measures IRN's PS, and that harmonic sounds elicit a strong impression of pitch, one might consider using HarmonicRatio to measure PS in the general case. However, Appendix \ref{ref:appendix2} suggests that this measure will be at least partially inaccurate.


\vspace{1cm}
\section{Variable pitch strength in contemporary popular music}\label{sec:variable}

\paragraph{Entire songs - examples.} An entire song or extract can exhibit a higher PS than another. For example, Vitalic's `Eternity' \citep{vitalic2017eternity} has a higher PS than Vitalic's `And it goes like' \citep{vitalic2023anditgoeslike}. Using \citepos{zwicker1990pitchstrength} sounds as a reference (see Appendix~\ref{ref:appendix3}), the beginning of `Eternity' shows a PS greater than sound 3 (between sounds 1 and 3), whereas the beginning of `And it goes like' shows a PS comparable to sounds 6 to 8.

\paragraph{Pitch strength evaluation.} How to evaluate the extent to which PS varies across many songs? The proposed answer involves (1) a \emph{noisiness-inharmonicity} space (see Appendix~\ref{ref:appendix4}) and (2) the Resonance EQ audio plug-in (see Appendix~\ref{ref:appendix5}):

\begin{enumerate}[label={(\arabic*)}]

\item The two dimensions of the noisiness-inharmonicity space relate to PS:
\begin{itemize}[noitemsep]
\item Dimension 1 is a measure of peak salience. More salient peaks can be associated with a higher PS: given a mixture of a harmonic sound and white noise, as the salience of the harmonic sound's partials increases, so does PS.\footnote{See \url{https://youtu.be/fUo7Toau8Xs}} This observation aligns with \citet{zwicker1990pitchstrength}, who noted, as mentioned above, that (a) sounds with line spectra typically generate a strong sense of pitch, while those with continuous spectra produce a weaker pitch sensation, and (b) the PS of pure tones (in this case, forming a harmonic complex tone) can be significantly reduced by partial masking sounds (in this case, the noise).
\item Dimension 2 is derived from HarmonicRatio (AC1), previously associated with PS in the case of IRN \citep{patterson2000perceptual}.
\end{itemize}

\item Resonance EQ's `gain' parameter modifies peak salience and therefore modifies PS.\footnote{See \url{https://youtu.be/yE8NfA82KhU}}

\end{enumerate}

\begin{figure}[h!]
  \centering
  \includegraphics[width=.65\columnwidth]{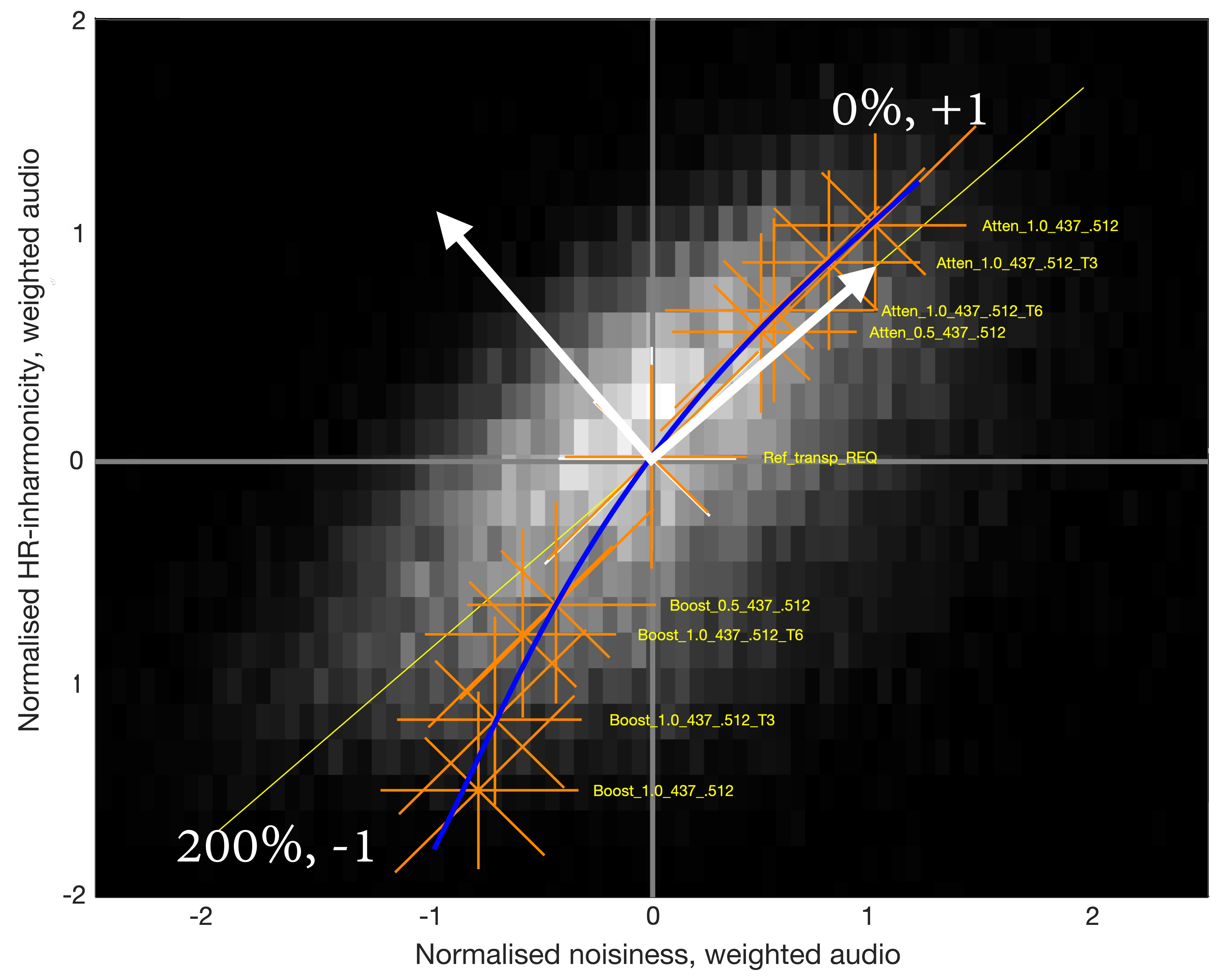}
  \caption{White arrows, PC1 and PC2 for the original distribution. Orange crosses, dataset processed with Resonance EQ for a variety of settings, 25\textsuperscript{th} and 75\textsuperscript{th} percentiles in each direction. Blue line, polynomial approximation of the 50\textsuperscript{th} percentiles.}
\label{fig:PC1PC2}
\end{figure}

The BEA dataset (see Appendix~\ref{ref:appendix6}) is processed with Resonance EQ (-1 to +1, or 0 to 200\%). As illustrated in \textbf{Figure~\ref{fig:PC1PC2}}, the processed noisiness-inharmonicity values cover a range similar to the original dataset's.

\textbf{Figure~\ref{fig:resonanceEQ}} in Appendix~\ref{ref:appendix5} shows how this span translates in terms of the power spectrum. The wide range of smoothness covered by the processed dataset suggests that the corresponding audio covers a wide range of PS.

\begin{figure}[h!]
  \centering
  \includegraphics[width=.65\columnwidth]{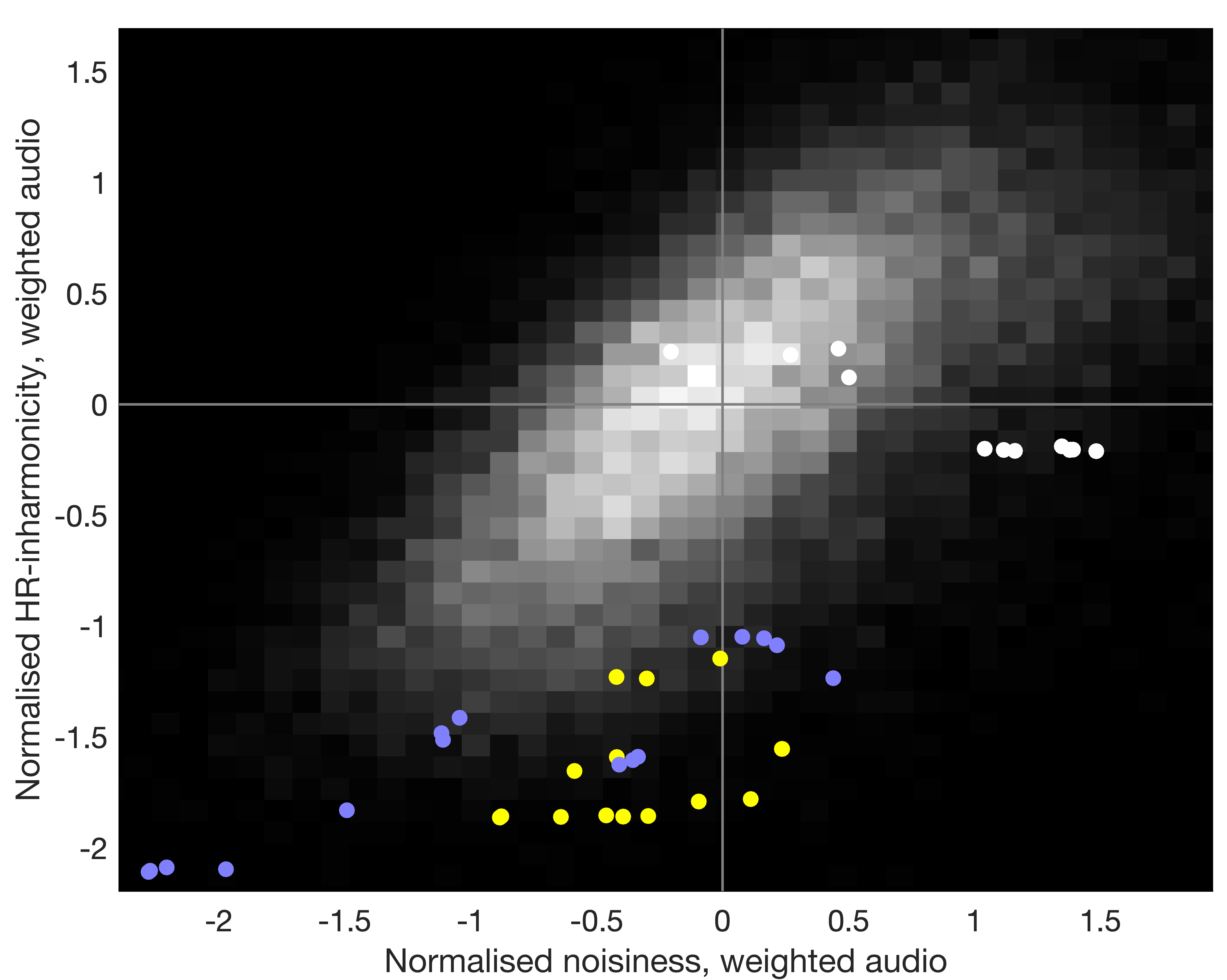}
  \caption{\citet{primaal2023boom}, positions of each beat in the noisiness-inharmonicity space. White, drum stem. Yellow, bass stem. Blue, vocal stem.}
\label{fig:PC1PC2hypermusic}
\end{figure}

\paragraph{Tracks within a song.} Individual tracks within a song can exhibit different PS. In \citet{primaal2023boom}, the vocal stem has a higher PS than the bass stem, which in turn has a higher PS than the drum stem. Using \citepos{zwicker1990pitchstrength} sounds as a reference, the vocal stem compares to sounds 4--5, the bass stem to sounds 6--7, and the drum stem to sounds 8--9. \textbf{Figure~\ref{fig:PC1PC2hypermusic}} shows the positions of the stems' beats in the noisiness-inharmonicity space.

\paragraph{Low PS does not prevent melodies.} In trap music, low PS melodies may be found in hi-hat tracks. The transcription in \citet[video 1]{suicideboys2015clouds} features such a track. Using \citepos{zwicker1990pitchstrength} sounds as a reference, the hi-hats compare to sounds 7--9 in terms of PS.

\paragraph{Summary.} In summary, contemporary popular music involves highly variable PS elements.

\clearpage

\clearpage

\newpage

\vspace{1cm}
\section{Pitch strength and structure}\label{ref:structure}

\paragraph{Large- and small-scale structure.} \citet[p~.13]{deruty2013methodological} summarize the distinction between large-scale and small-scale musical structures.

Large-scale structure, `semiotic structure', or `sectional form' \citep{bimbot2012semiotic}, describes the long-term regularities in a musical piece \citep{bimbot2014semiotic}. Segments in this structure are `autonomous and comparable blocks' \citep{bimbot2010decomposition,bimbot2014semiotic}, with a median duration of about 11.5 seconds \citep{deruty2013methodological,snyder2000music}. In Western classical music, these segments may be classified as `sentences' and `periods' \citep{caplin1998classical,schoenberg1967fundamentals}, and their relations can be described through comparison \citep{bimbot2012semiotic}.

Small-scale structure refers to the internal organization of these segments, termed `form' by \citet{caplin1998classical}. This time scale is known as the `morpho-syntagmatic level' \citep{bimbot2014semiotic,bimbot2016system} or the level of `melodic and rhythmic grouping' \citep{monelle2014linguistics}. Functions at this level include antecedent, consequent, basic idea, contrasting idea \citep{caplin1998classical}, and contrast \citep{bimbot2016system}. Relationships between units are described through implication \citep{bimbot2016system}, with the contrast identified by deviations from implied relations.

\begin{figure}[htbp]
  \centering
  \includegraphics[width=1\columnwidth]{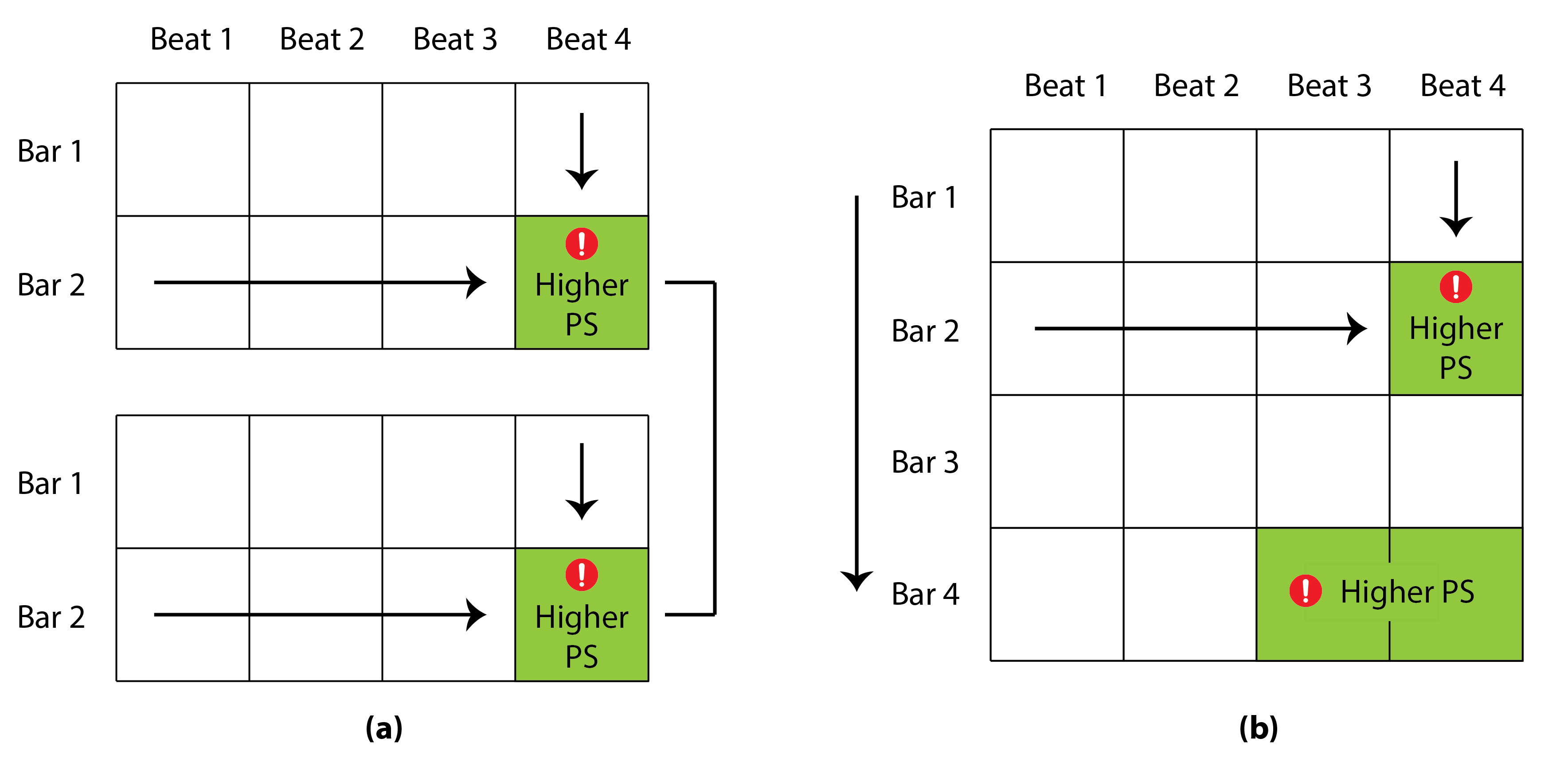}
  \caption{Some implications and contrasts involving PS in (a) \citet{primaal2023cardinal} and (b) \citet{primaal2023whomp}. The arrows denote implications, the angled black line denotes a comparison, and the colored square show the contrasts.}
\label{fig:implications}
\end{figure}

\paragraph{Small-scale structure and PS, examples.} \textbf{Figure \ref{fig:implications}(a)} shows two segments from \citet{primaal2023cardinal}. In the first segment, the eighth beat has a higher PS, with all previous tracks muted and replaced by a synthesizer with an evolving resonant filter. In the second segment, the eighth beat has an even higher PS, with all previous tracks muted and replaced by choirs.

\textbf{Figure \ref{fig:implications}(b)} shows a segment from \citet{primaal2023whomp}. The 16 beats are grouped into one segment due to a faster tempo than \citet{primaal2023cardinal}. A contrast on the eighth beat has a higher PS, featuring a loud vocal part absent in the previous beats. Another contrast occurs at the end of the fourth bar, where all previous tracks are muted and replaced by a tonal instrumental loop.

\begin{figure}[htbp]
  \centering
  \includegraphics[width=1\columnwidth]{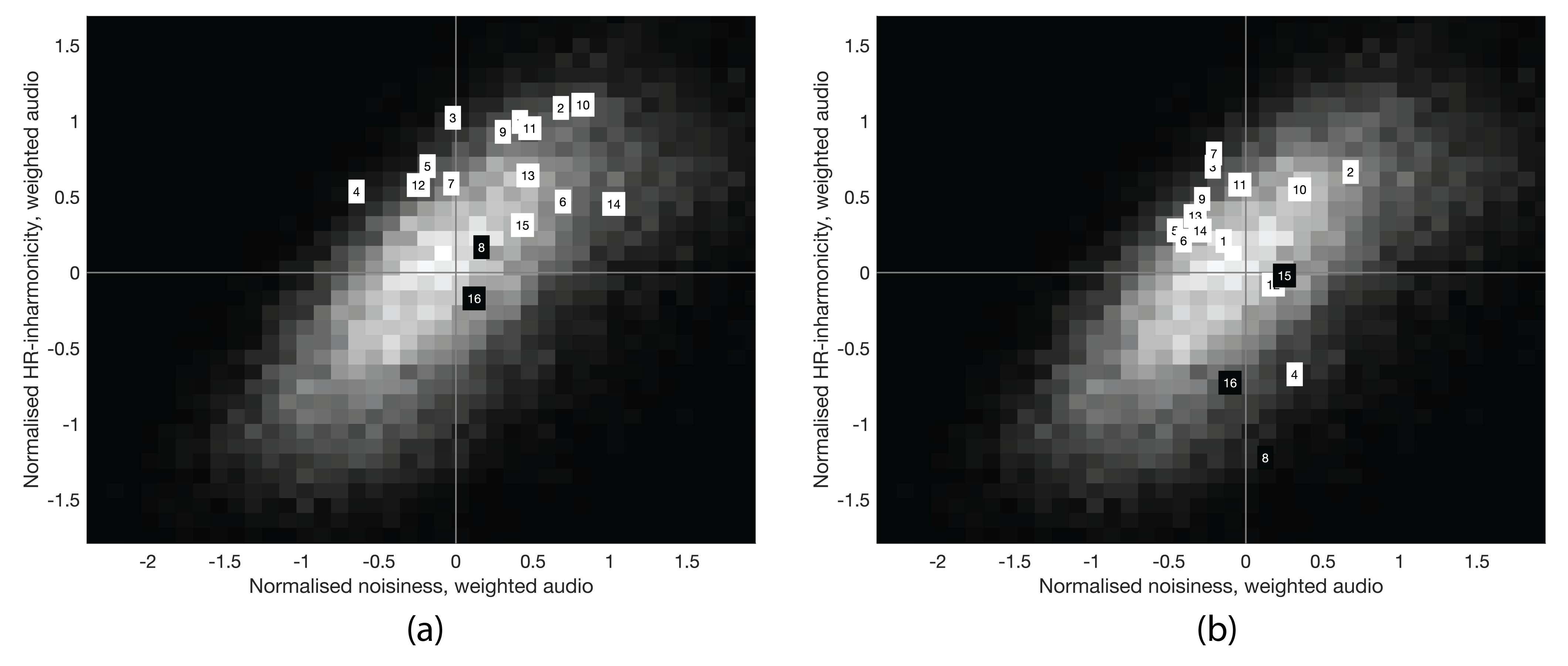}
  \caption{Position of each beat in the noisiness-inharmonicity space. (a) \citet{primaal2023cardinal} and (b) \citet{primaal2023whomp}.}
\label{fig:PC1PC2positions}
\end{figure}

\textbf{Figure \ref{fig:PC1PC2positions}} shows the positions of the beats from \citet{primaal2023cardinal,primaal2023whomp} in the noisiness-inharmonicity space. Beats 8 and 16 are closer to the bottom left than the others.

The implications and contrasts shown in \textbf{Figure \ref{fig:implications}} indicates that in \citet{primaal2023cardinal,primaal2023whomp}, PS is part of the determinant of form \citep{caplin1998classical}, contrasting with the music of Mozart, Haydn, and Beethoven according to the same author.

\begin{figure}[htbp]
  \centering
  \includegraphics[width=1\columnwidth]{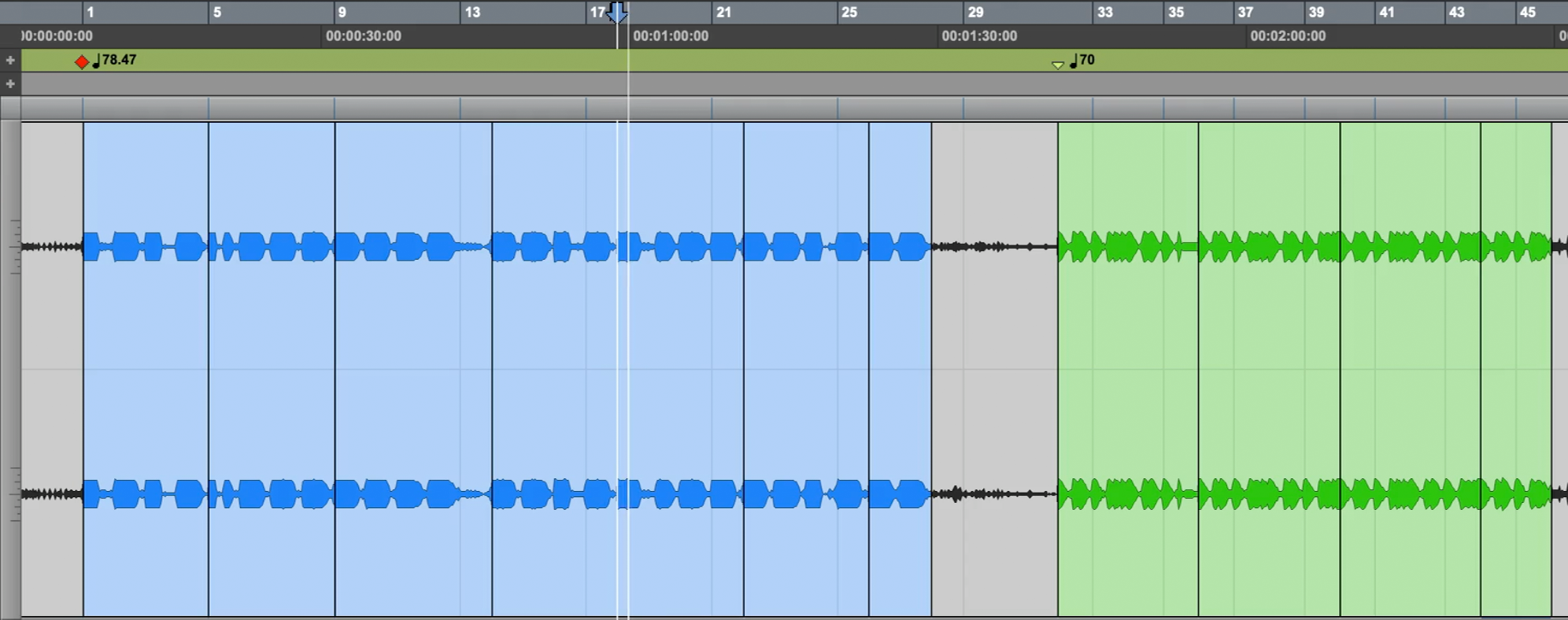}
  \caption{Decomposition of \citet{jid2022raydar} into semiotic segments. The segments in green have a higher pitch strength than the segments in blue.}
\label{fig:JID}
\end{figure}

\paragraph{Large-scale structure and PS, examples.} In terms of large-scale structure, PS can help differentiate between semiotic segments. In \citet{primaal2023yadayada}, the segment between 0:46 and 0:58 has lower pitch strength than the segment between 0:58 and 1:11 due to a more continuous bass and a louder, more stable flute part. \textbf{Figure~\ref{fig:JID}} shows the decomposition of \citet{jid2022raydar} into semiotic segments. The song features seven segments with relatively low PS, where a rap flow is delivered over a low TR-808 kick drum. After a break, the song transitions into four segments with higher PS, where the flow is closer to being sung, and a keyboard chord is repeated.

\paragraph{Summary.} In summary, PS is a musical parameter in contemporary popular music that may be involved in both small- and large-scale structures.


\newpage
\section{Pitch strength and dissonance}\label{ref:dissonance}

\paragraph{Low PS can help manage dissonance.}  Following \citet{tenney1988consonance}, \citet[pp.~77--81]{sethares2005tuning} distinguishes between five types of consonance-dissonance, numbered CDC-1 to CDC-5. The dissonance interpretation here primarily relates to CDC-2 (polyphonic consonance/dissonance).

\paragraph{Drums and pitch.} Drums can contain pitched content \citep{richardson2010acoustic}. Drum eigenfrequencies are not necessarily harmonic but can be made harmonic under certain conditions \citep{antunes2017possible}. Usually, the tonal elements in drums are not structured like partials in a harmonic series. Their frequency relationship ranges from inharmonic to chaotic \citep{wu2018review}. 

\begin{figure}[htb]
  \centering
  \includegraphics[width=1\columnwidth]{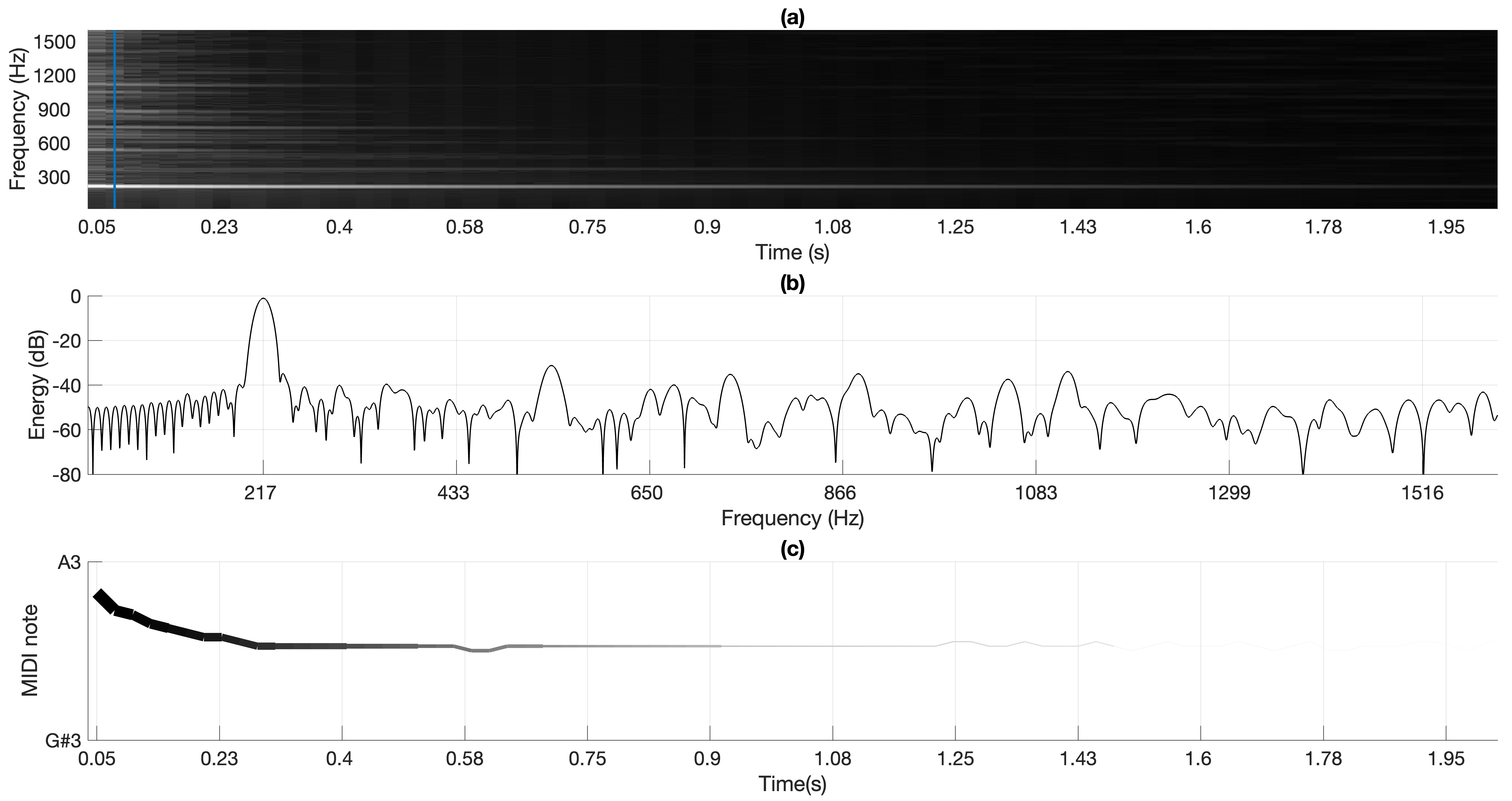}
  \caption{Sample `B-A3 2.wav' (A3, medium velocity) from the sample bank `Royotoms by Alan ViSTa'. (a) ISO 226--2003 weighted sample, short-term Fourier transform. (b) Power spectrum corresponding to the frame highlighted in blue in (a). The grid is set on the multiples of the lowest partial. (c) MIDI note corresponding to the peak of maximum energy in (a).}
\label{fig:rototoms}
\end{figure}

Drums can be tuned. In popular music, the tuning of acoustic drums can have a significant impact on the quality and contextuality of the instrument \citep{toulson2009perception}. Some drums, such as the timpani \citep{blades2001timpani} and roto-toms \citep{holland2001rototoms}, have a stronger pitch than others. \textbf{Figure~\ref{fig:rototoms}} shows the signal properties of a medium-velocity, middle-range roto-tom sample. The sample has a weak pitch: its energy decreases rapidly, features inharmonic upper partials, and is noisy. The partial with the highest energy is 40dB above the noise floor, and other partials are, at best, 10dB above it. In terms of perceived pitch strength, the sample is comparable to \citepos{zwicker1990pitchstrength} sound 7.

\paragraph{Low PS to handle dissonance: examples.} \citet{pinkfloyd1973time} features an introduction with prominent roto-toms. One version \citep[video 2]{rogerwaters2020time} includes a clear recording of the roto-tom player. In Roger Waters \citep[video 2; video 3, 1'05]{rogerwaters2020time}, the roto-toms play natural Gs in F\# minor. The low PS of the roto-toms attenuates the dissonance, making it acceptable for the music genre. In video 3, Resonance EQ increases PS, making the roto-toms easier to transcribe and dissonances more obvious. Video 4 shows how transcription with a high-PS instrument (a piano) induces stronger dissonances.

\newpage

Similar examples appear in rap music, where lead vocals resemble \textit{Sprechgesang} \citep{griffiths2001sprechgesang} with low PS. In \citet[video 1]{eminem2010nolove}, the song is in D\# minor, and the line `Scribble out them rhymes you were gonna spit' (3'21) features several instances of D-flat. As with \citet{rogerwaters2020time}, dissonance intensifies when processed with Resonance EQ \citep[video 2]{eminem2010nolove}.

\paragraph{Low PS and non-transposition of out-of-tune samples.} The Hyper Music producers (see Appendix~\ref{ref:appendix7}) note that when a sound is out of tune but has low PS, they often avoid transposing it to preserve its timbre. The low PS reduces dissonance. This practice aligns with \citet[p.~82]{frisius2010search}, who states that a music theory positing neutral transposition -- where pitch groups retain their character when transposed -- overlooks the transposition of the sounds themselves. \citeauthor{frisius2010search} cites the composer Luigi Russolo, who found it challenging to transpose sonic gestures into other registers without losing their identity, particularly when the pitch is not clearly definable.

\paragraph{Summary.} Low PS elements provide greater flexibility in using polyphonic dissonance in contemporary popular music.


\newpage
\section{Low-pitch strength audible harmonics}\label{ref:harmonics}

\paragraph{Examples.} A previous section cited \citet{mybloodyvalentine1991sometimes}, in which the upper harmonics are loud enough to become audible. A key aspect of Hyper Music producers' work is amplifying upper partials while attenuating the fundamental, making the upper partials audible as carriers of perceivable pitches. For example, in the bass part from \citet[video 2]{primaal2023cardinal}, harmonic 5 (two octaves and a major third) is boosted using equalization. The boosted harmonics' pitch isn't necessarily strong. Producers use these harmonics' low PS to create ambiguity in the perceived number of pitches.

\paragraph{Harmonic tones may convey multiple pitches.} Perceptual tests asked respondents how many simultaneous pitches they could hear in two bass tracks and a keyboard track \citep{primaal2023elevatesilvershomp}. The boosted harmonics were perceived as independent pitches, with the number varying from 1 to 4. The mean number of perceived pitches was 1.7, 1.95, and 2.45, respectively.

\paragraph{Role of distortion.} Hyper Music producers may use distortion to boost or generate harmonics. Distortion applies a non-linear transformation to the signal, causing intermodulation distortion \citep[p.~464]{newell2017recording}, which produces partials from each input partial and their combinations. If $f_a$ and $f_b$ are the input frequencies, the output contains frequencies $k_a f_a + k_b f_b$, where $(k_a, k_b) \in \mathbb{Z}^2$. Distorting a sine wave generates harmonics, while distorting a harmonic complex tone boosts harmonics, making them audible. Thus, distortion adds low-PS elements.

In \citet{massiveattack2010saturday}, the two guitars and keyboard are distorted, forming streams in the sense of \citet{bregman1994auditory}. Each stream has pitches of varying strength, with harmonics' pitches being less audible than the perceived fundamental.

\paragraph{Summary.} Using equalization or distortion to boost upper harmonics so that they become low-PS audible components is a way to add perceptual richness to the music.

\vspace{.5cm}


\section{Conclusion and future work}\label{ref:conclusion}

\paragraph{Summary.} The chapter presents arguments for the importance of PS in contemporary popular music. Signal analysis, combined with perceptual observations, suggests several conclusions:

\begin{itemize}[noitemsep]
	\item Peak salience, and therefore, PS, vary significantly across songs. 
	\item Deviations to small-scale form implications may involve PS, making it part of contemporary popular music's determinant of form.
	\item PS may also contribute to the large-scale form.
	\item Polyphonic dissonance may be attenuated if at least one involved element has low PS.
	\item Amplified upper harmonics may be individually perceivable elements with low PS.
\end{itemize}

The notion of PS may, therefore, improve our understanding of contemporary popular music. It may also be a useful addition to generative AI models to improve their alignment with human perception and musical practice.

\paragraph{Perspectives.} Future work should aim at consolidating the chapter's content. The chapter suggests, but doesn't prove, the importance of PS in contemporary popular music. It also suggests, but doesn't demonstrate, the concept's usefulness in generative AI models. It introduces relations between PS and the noisiness-inharmonicity space but doesn't formally link one to the other. Also, most perceptual observations it contains are the author's; tests should be conducted so that such observations can be made more reliable.

\vspace{2cm}

\appendix
\section*{APPENDICES}

\vspace{.1cm}
\section{ISO226:2003 weighting}\label{ref:appendix1}

Humans are not equally sensitive to all frequencies \citep{fletcher1933loudness}. Models, such as \citet{iso2262003}, describe how tones are perceived as equally loud depending on their frequency and sound pressure level. This chapter uses the ISO 226-2003 weighting to ensure that evaluated power spectra more accurately reflect the listener's perceived spectral profile.



\vspace{.1cm}
\section{Pitch strength does not correlate with HarmonicRatio in the general case}\label{ref:appendix2}

\paragraph{Argument 1.} The first argument suggesting that pitch strength doesn't correlate with HarmonicRatio (AC1) in the general case derives from the measurement of HarmonicRatio for the 11 sounds from \citeauthor{zwicker1990pitchstrength} (see Appendix~\ref{ref:appendix1}). \textbf{Figure \ref{fig:harmonicratios}(a)} shows HarmonicRatio for the 11 sounds, and \textbf{Figure \ref{fig:harmonicratios}(b)} shows 10\textsuperscript{HarmonicRatio} for the same sounds. HarmonicRatio is close to one for the first seven sounds and much lower for the last four.

Comparing \textbf{Figure~\ref{fig:harmonicratios}(a)(b)} with Figures 5.25(a--c) from \citet[p.~137]{zwicker1990pitchstrength}, in which the PS values decrease regularly, suggests a potential correlation between PS and HarmonicRatio, though the scale might have to be adjusted. To adjust its scale to actual music, HarmonicRatio is measured on the BEA dataset (see Appendix~\ref{ref:appendix5}), and the resulting distribution is normalized using the transformation (1-HarmonicRatio)\textsuperscript{0.21}. \textbf{Figure~\ref{fig:harmonicratios}(c)} shows the transformed HarmonicRatio for the 11 sounds. The resulting values still show little correlation with the PS measures by \citet{zwicker1990pitchstrength}. As a result, it is not possible to conclusively link HarmonicRatio and pitch strength in the case of the 11 sounds provided by \citet{zwicker1990pitchstrength}.

\begin{figure}[htb]
  \centering
  \includegraphics[width=1\columnwidth]{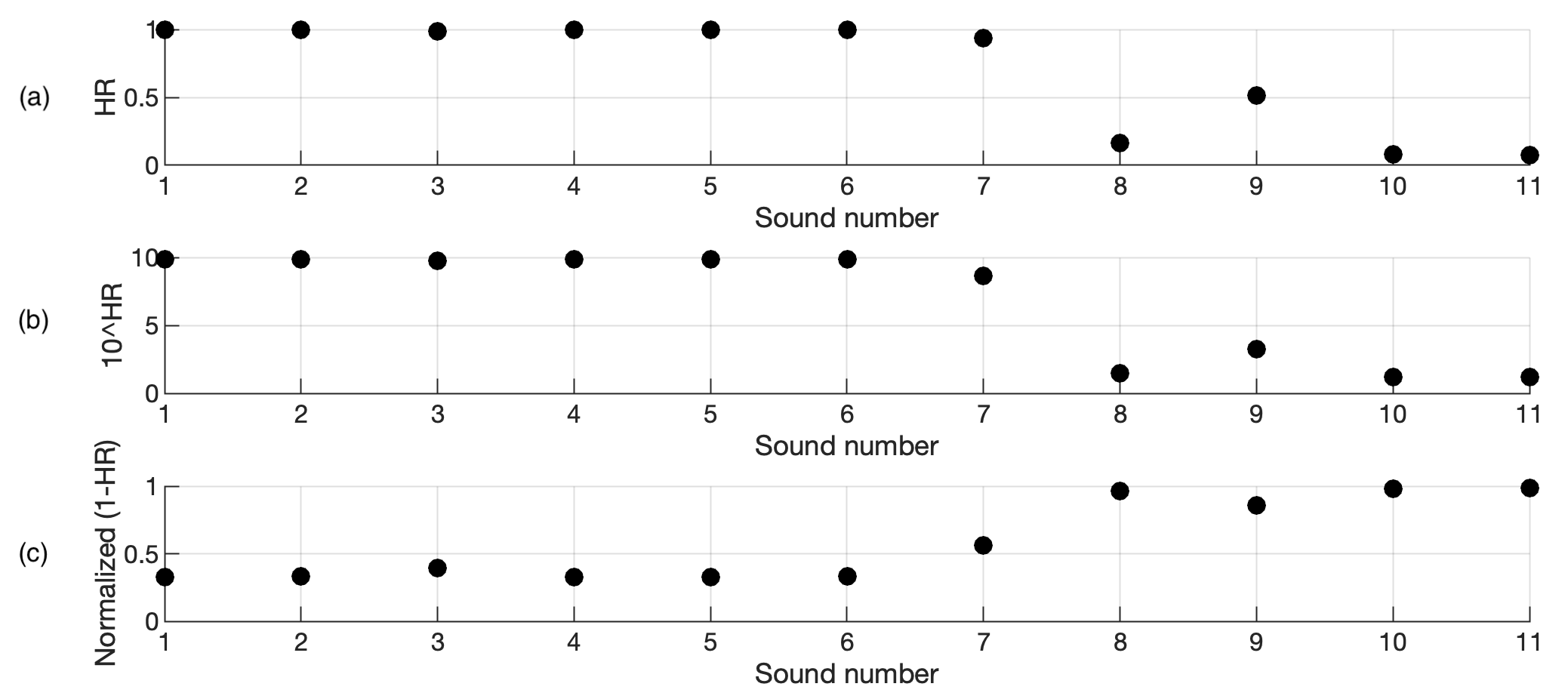}
  \caption{(a) HarmonicRatio for the 11 sounds used by \citet[p.~136]{zwicker1990pitchstrength}. (b) $10^{HarmonicRatio}$ for the same sounds. (c) $(1-HarmonicRatio)^{0.21}$ for the same sounds.}
\label{fig:harmonicratios}
\end{figure}

\newpage

\paragraph{Argument 2.} The second argument stems from music typically containing multiple simultaneous tones. Here, HarmonicRatio is compared for two pairs of harmonic complex tones: a minor third and a fourth, both in equal temperament. The tones are generated using \citepos{mauch2010approximate} model, where the amplitude of the $k$th partial is $s^{k-1}$ with $s=0.8$ and 10 harmonics. On a scale from 0 to 1, the HarmonicRatio for the minor third is 0.829 and 0.964 for the fourth. After normalization, these values become 0.690 and 0.497. Despite this, the pitch strengths for both sounds remain similar.

\begin{figure}[htb]
  \centering
  \includegraphics[width=1\columnwidth]{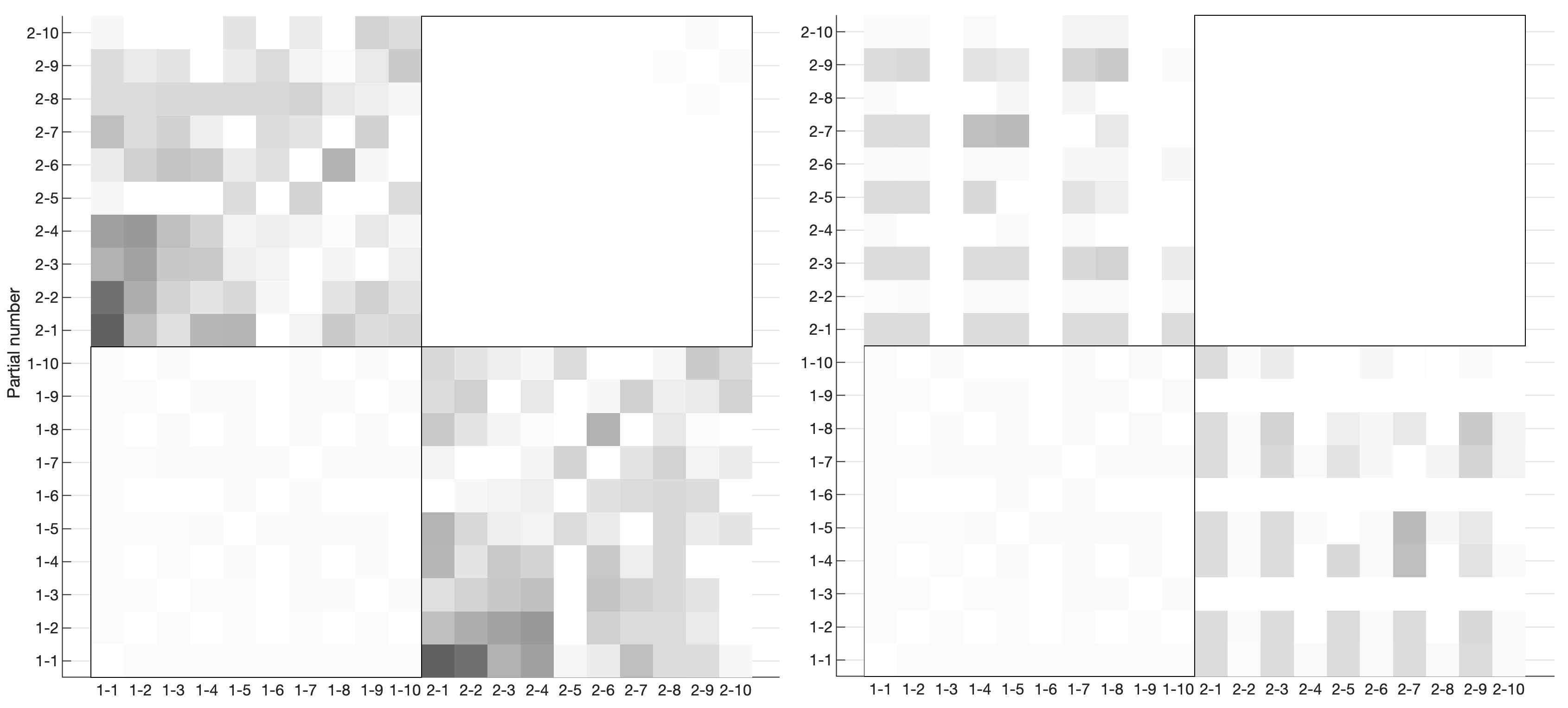}
  \caption{HarmonicRatio derived from the pairwise combination of elements in two harmonic tones. Clearer shades correspond to higher values, and darker shades to lower values. The interval between the two tones is three half-tones (left) and five half-tones (right). Partial numbers are written as `tone number-partial number' (e.g., `2--3' corresponds to the third partial of the second tone).}
\label{fig:pairwise}
\end{figure}

As shown in \textbf{Figure~\ref{fig:pairwise}}, the difference in HarmonicRatio between the fourth and minor third can be explained by evaluating pairwise HarmonicRatios for each harmonic. The intervals in the fourth are generally purer than those in the minor third.

\paragraph{Summary} These arguments suggest that while the PS of an IRN wave may relate to the height of the first peak in its normalized autocorrelation function (AC1), this observation is not universal.

\section{\citeauthor{zwicker1990pitchstrength}'s reference sounds}\label{ref:appendix3}

\citet[p.~136]{zwicker1990pitchstrength} measured the PS of 11 sounds:

\begin{enumerate}[noitemsep]
	\item Pure tone.
	\item Harmonic complex tone, -3dB/octave, low-pass filter (only 7 partials).
	\item Harmonic complex tone, -3dB/octave.
	\item Narrow-band noise,  \(\Delta \)f = 10Hz.
	\item AM tone, modulation index m=1.
	\item Harmonic complex tone, band-pass.
	\item Band-pass noise, filter slope 96dB/octave.
	\item Low-pass noise, filter slope 192dB/octave.
	\item Comb-filter noise, attenuation d=40dB.
	\item AM noise, modulation index m = 1.
	\item High-pass noise, filter slope 192dB/octave.
\end{enumerate}

Each sound is available at three frequencies: 125, 250, and 500Hz. \citeauthor{zwicker1990pitchstrength} numbered the sounds in order of decreasing PS, which diminishes relatively consistently (p. 137).

In the chapter, examples are compared in terms of PS to the 500Hz version of these 11 sounds. Since \citeauthor{zwicker1990pitchstrength}'s measures are normalized for each sound frequency, PS is referenced by sound number rather than an absolute unit.


\begin{figure}[h!]
  \centering
  \includegraphics[width=1\columnwidth]{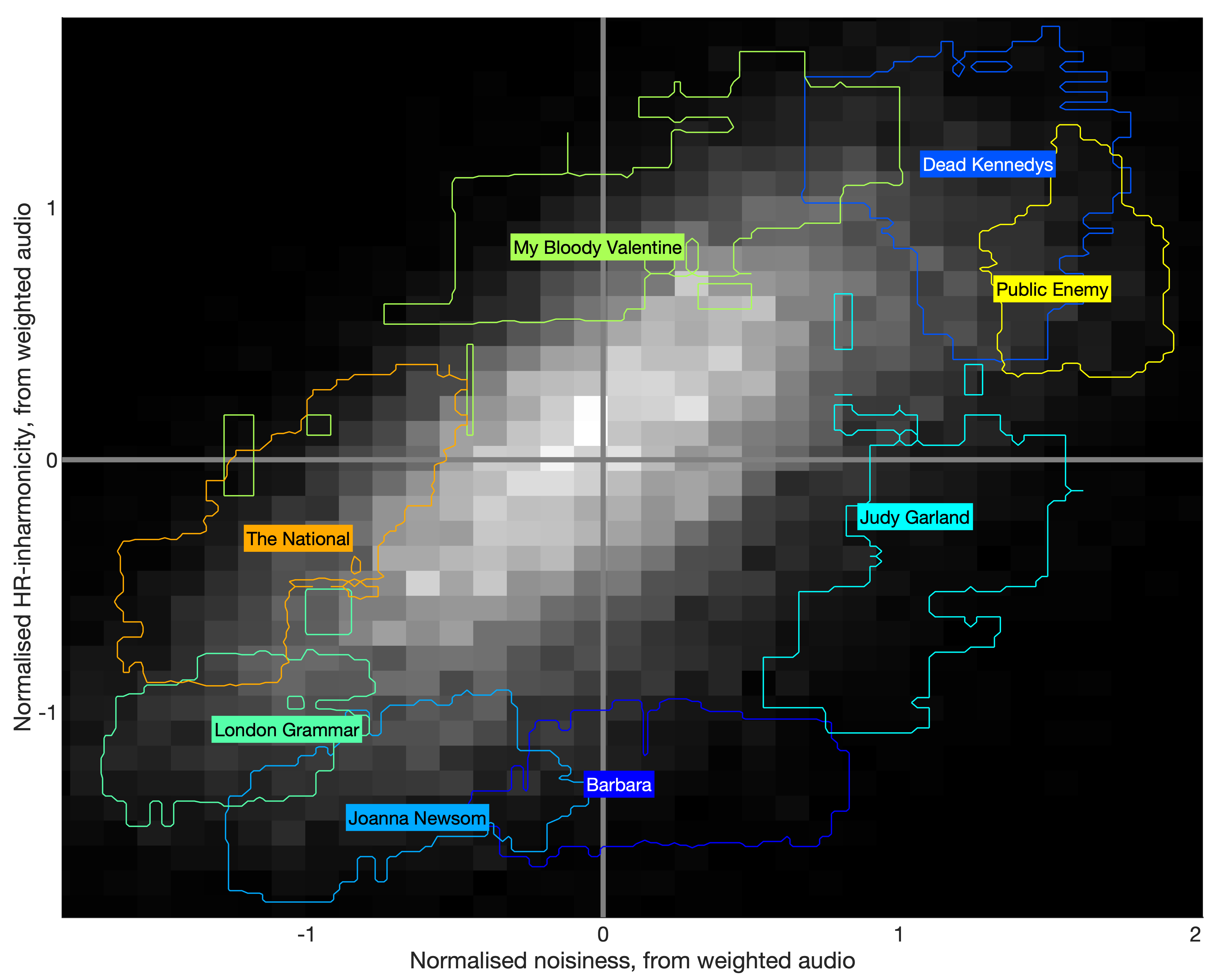}
  \caption{BEA dataset, distribution of normalised noisiness and HR-inharmonicity. For each featured artist, the contours represent the isolines at half-height.}
\label{fig:space}
\end{figure}

\section{The noisiness-inharmonicity space}\label{ref:appendix4}

\citet{deruty2024inharmonicity} derive a noisiness-inharmonicity space from spectral flatness \citep[p.~29]{isompeg7audio}\citep{peeters2004large} and HarmonicRatio \citep[p.~36]{isompeg7audio} evaluated on the BEA dataset (see Appendix \ref{ref:appendix6}). As illustrated in \textbf{Figure~\ref{fig:space}}, the space provides a coherent representation in which the distribution's principal components are normal and where songs from one artist may be found next to each other.


\section{Resonance EQ}\label{ref:appendix5}

\begin{figure}[htbp]
  \centering
  \includegraphics[width=1\columnwidth]{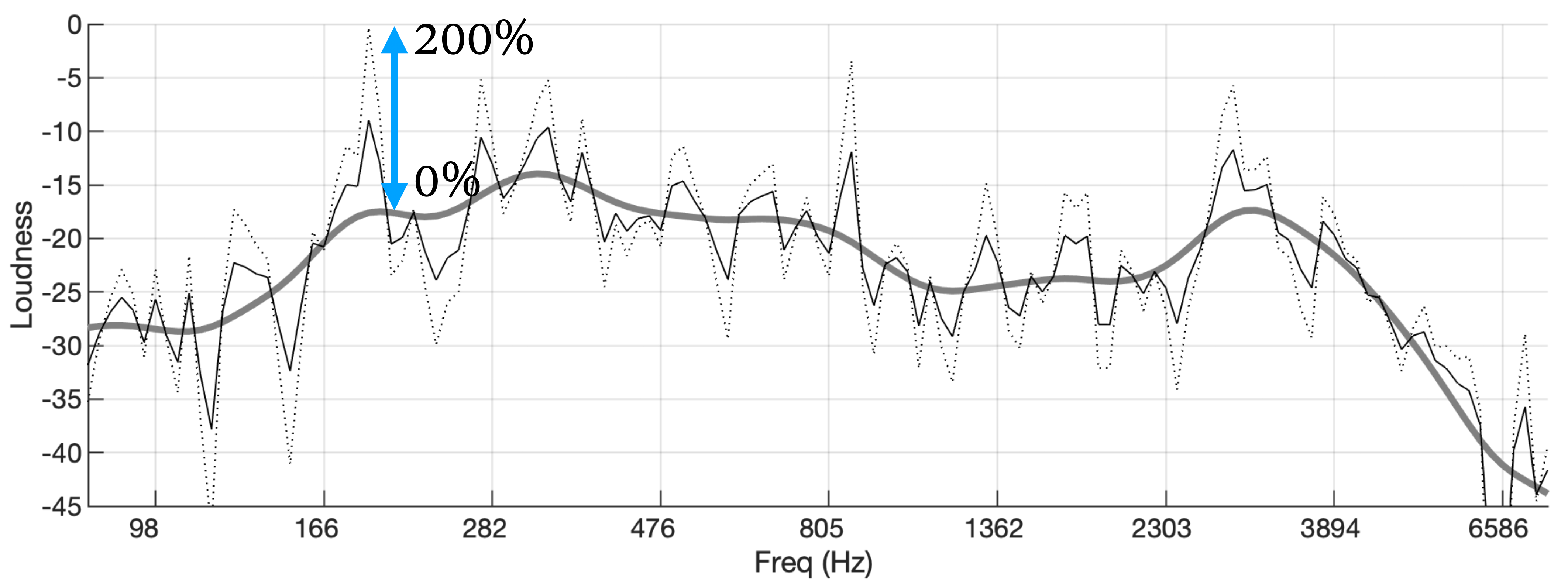}
  \caption{Solid black line, original power spectrum. Solid gray line, power spectrum with Resonance EQ's gain at -1. Dotted line, power spectrum with Resonance EQ's gain at +1.}
\label{fig:resonanceEQ}
\end{figure}

The chapter uses Resonance EQ, a VST/AU plug-in from Sony Computer Science Laboratories.\footnote{\url{https://cslmusicteam.sony.fr/prototypes/resonance-eq/}} The plug-in modifies the salience of spectral peaks using a `gain' parameter. As shown in \textbf{Figure~\ref{fig:resonanceEQ}}, at zero gain, spectral peaks are unaffected. At +1, formant salience is doubled. At -1, salience is nullified.

\newpage
\section{The BEA dataset}\label{ref:appendix6}

The BEA dataset contains 30,435 tracks released between 1961 and 2022, with 464 to 556 tracks per year. It extends the dataset used by \citet{deruty2015mir}. Tracks are chosen from the `Best Ever Albums' website,\footnote{\url{www.besteveralbums.com}} a review aggregator that lists the best-rated albums each year.


\section{Hyper-Music}\label{ref:appendix7}

Hyper Music, mentioned in \citet{deruty2022development,deruty2022melatonin}, is a sound and music production company that creates music for ads, TV series, and feature films. They have a catalogue of music tracks that they submit to potential clients when requested. If a track is selected, Hyper Music customizes it for the client, acting as a provider of production music \citep{Productionmusic}. This work uses music examples from Hyper Music and conclusions from informal interviews with their producers.

\clearpage

\bibliographystyle{apalike}
\bibliography{mybib.bib}
\end{document}